\documentstyle[epsfig]{aipproc}
\begin{document}
\title{Single production of charged Higgs bosons 
       at linear colliders \footnote{Talk given by S.~Kanemura
       at {\it the Linear Collider Workshop 2000, October 24-28, 
       2000, Fermilab, USA.}}}

\author{S.~Kanemura$^{\; a}$, S.~Moretti$^{\; b}$ and 
K.~Odagiri$\; ^c$}
\address{$^{a}$Physics and Astronomy Department, 
               Michigan State University,
               East Lansing, MI 48824-1116, USA\\
$^b$CERN, CH 1211 Geneva 23 Switzerland\\
$^{c}$Theory Group, KEK, 1-1 Oho, Tsukuba, Ibaraki 305-0801, Japan}

\begin{flushright}
CERN-TH/2001-025\\
KEK-TH-743\\
MSUHEP-01130
\end{flushright}

\maketitle

\begin{abstract}
We discuss single charged Higgs boson production from $e^+e^-$ 
annihilation at next generation linear colliders. They can be 
important to study the phenomenology of charged Higgs bosons, 
especially when charged Higgs bosons are too heavy to be produced by 
the pair production mechanism. Cross sections for various single 
production processes are evaluated at the leading order. 
Our analysis shows that in some parameter regions the phenomenology 
of charged Higgs bosons can be explored even beyond the kinematic 
limit for pair production by using single production processes. 
\end{abstract}
 
\section*{Introduction}

Electroweak symmetry breaking sectors which are composed of 
two scalar isospin-doublets predict charged Higgs bosons ($H^\pm$). 
Their discovery at future colliders will directly indicate such a 
non-minimal structure of the Higgs sector, and its detailed 
information will give strong hints for the structre of physics 
beyond the Standard Model (SM). Therefore, their phenomenology 
has been evaluated especially over the past few years 
as their discovery potential has become more clear.

At CERN Large Hadron Collider, charged Higgs bosons below the top 
quark mass will be produced in the top quark decay process. 
For heavier charged Higgs bosons, they will be produced mainly through 
the subprocesses $gb\to tH^-$, $gg\to t\bar bH^-$ and their charge 
conjugate, and they may be probed, for example, using the decay modes 
$H^-\to\tau^-\bar\nu$ and its charge conjugate especially for large 
$\tan\beta$ values, in spite of huge QCD backgrounds.

On the other hand, at future electron-positron linear colliders (LC's), 
the dominant production process should be the pair 
production\cite{ee_pair} $e^-e^+ \to H^+H^-$. Its cross section  
depends only on the charged Higgs mass, and this process provides 
a hallmark channel through which we can study $H^\pm$ 
phenomenology\cite{pair_phenom} as long as the charged Higgs boson mass 
($M^{}_{H^\pm}$) is sufficiently smaller than its kinematical threshold 
$\sqrt{s}/2$.  
When $M^{}_{H^\pm}$ is near or above $\sqrt{s}/2$, the pair production is 
not available. In such a case, the phenomenology of charged Higgs bosons 
may be explored through the kinematically-allowed single production 
processes.

%\section*{Single $H^\pm$ production processes}

In this talk, we discuss various single $H^\pm$ production channels 
to complement the pair production channel. 
We consider the following processes\cite{kmo}:
  \begin{eqnarray}
   e^-e^+ &\to& \tau^-\bar\nu_\tau H^+, \tau^+\nu_\tau H^-\label{proc_tau} \\
   e^-e^+ &\to& \bar tbH^+, t\bar bH^-              \label{proc_tb}  \\
   e^-e^+ &\to& W^\mp H^\pm \mathrm{(one\ loop)}    \label{proc_wh}  \\
   e^-e^+ &\to& e^-\bar\nu H^+, e^+\nu H^-
                \mathrm{ (one\ loop)}               \label{proc_enu} \\
   e^-e^+ &\to& Z^0W^\mp H^\pm                      \label{proc_zwh} \\
   e^-e^+ &\to& h^0W^\mp H^\pm                      \label{proc_lwh} \\
   e^-e^+ &\to& H^0W^\mp H^\pm                      \label{proc_bwh} \\
   e^-e^+ &\to& A^0W^\mp H^\pm                      \label{proc_awh} \\
   e^-e^+ &\to& e^-e^+W^\mp H^\pm                   \label{proc_eewh}\\
   e^-e^+ &\to& \nu_e\bar\nu_eW^\mp H^\pm           \label{proc_nnwh}\\
   e^-e^+ &\to& e^-\bar\nu_eZ^0 H^+, e^+\nu_eZ^0H^- \label{proc_enzh}\\
   e^-e^+ &\to& e^-\bar\nu_eh^0 H^+, e^+\nu_eh^0H^- \label{proc_enlh}\\
   e^-e^+ &\to& e^-\bar\nu_eH^0 H^+, e^+\nu_eH^0H^- \label{proc_enbh}\\
   e^-e^+ &\to& e^-\bar\nu_eA^0 H^+, e^+\nu_eA^0H^-.\label{proc_enah}
  \end{eqnarray}
The $\tau\nu_\tau$ associated production mode (\ref{proc_tau}) 
has been studied in the literature\cite{ee_taunuH}.
The processes (\ref{proc_tb}), (\ref{proc_lwh}), (\ref{proc_bwh}) 
and (\ref{proc_awh}), as well as (\ref{proc_zwh}) with $Z \to b\overline{b}$ 
lead to the same final state $b\bar b W^\pm H^\mp$\cite{ee_bbWH}.   
The one-loop induced processes $e^-e^+ \to H^\pm W^\mp$
have also been studied in Refs.\cite{hollik_HW,sk_HW}.
The one-loop induced $HWZ$ and $HW\gamma$ vertices 
which enter into (\ref{proc_enu}) 
have been calculated in Ref.\cite{rizzomendez}.

\section*{Results}

We evaluate cross sections of the above single production processes 
at the leading order. The Two-Higgs-Doublet-Model (2HDM) 
parameters are taken assuming the Minimal Supersymmetric Standard Model.  
For the SM parameters we adopted the following numbers:
$m_b=4.25$ GeV, $m_t=175$ GeV, $m_e=0.511$ MeV, $m_\tau=1.78$ GeV,
$m_\nu=0$, $M_W=80.23$ GeV, $\Gamma_W=2.08$ GeV, $M_Z=91.19$ GeV,
$\Gamma_Z=2.50$ GeV, $\sin^2\theta_W=0.232$. 
The Center-of-Mass energy of the collider ($\sqrt{s}$) is assumed to be 
$500$ GeV and $1000$ GeV. Details of the calculation 
and the plots are shown in Ref.\cite{kmo}. 

The $\tau\nu_\tau$ associated production processes (\ref{proc_tau}),  
shown in Fig.~1, are dominated by the pair production as long as 
it is kinematically allowed ($M^{}_{H^\pm} < \sqrt{s}/2$). 
Above this threshold,  
the cross sections still exceed $10^{-5}$ pb 
at large $\tan\beta$ values. If $H^\pm$ is heavy enough to decay into 
$tb$, dominant background contribution comes from top-quark pair 
production. 
By peforming $W$ and $t$ mass reconstruction, 
the background can be reduced naively by 
${\mathcal O}(\alpha^2_{\rm EW})$, so that the signal would be visible, 
although detailed simulations are needed depending on each machine. 

\begin{figure}[t]
  \epsfig{file=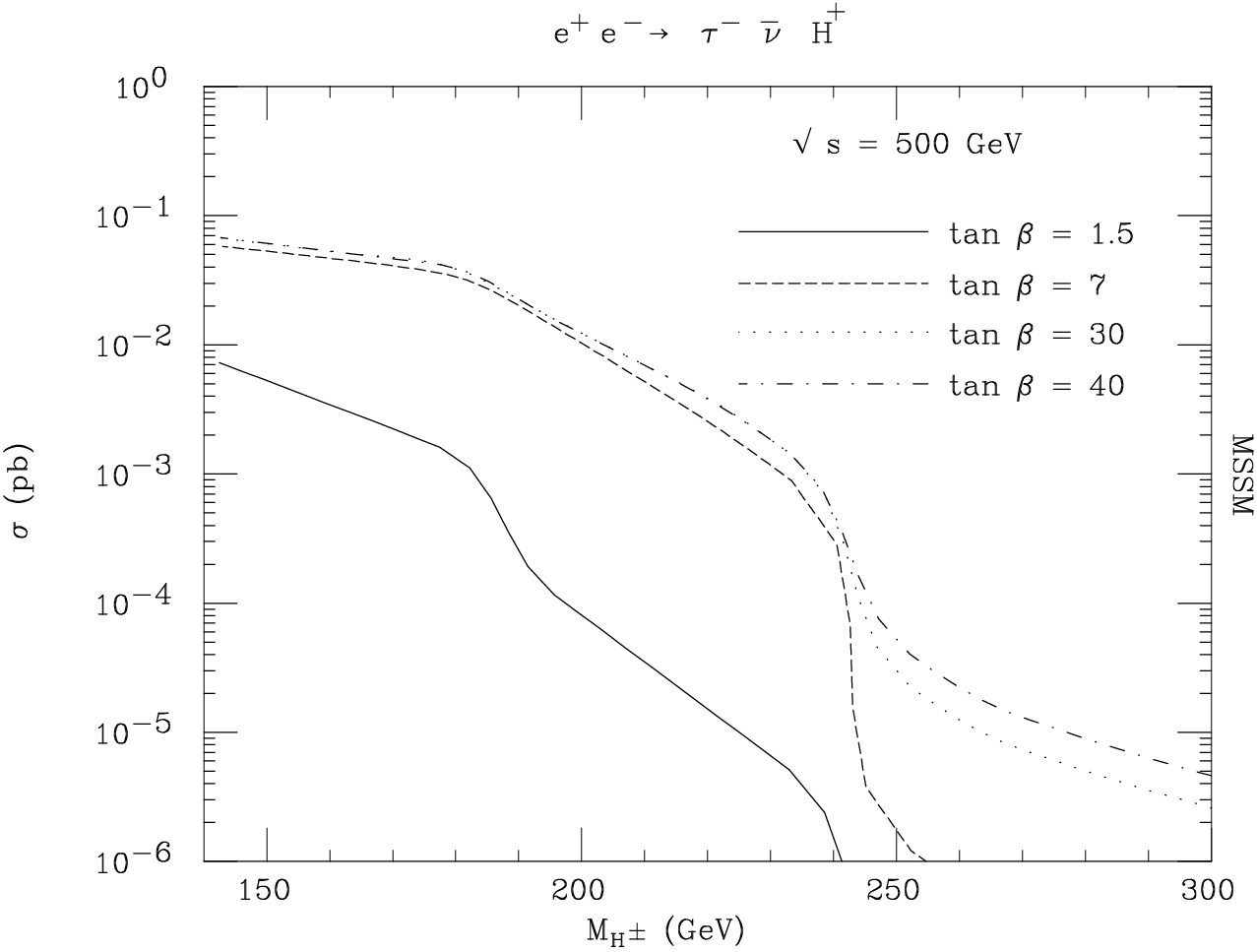,width=5cm,angle=90}
  \epsfig{file=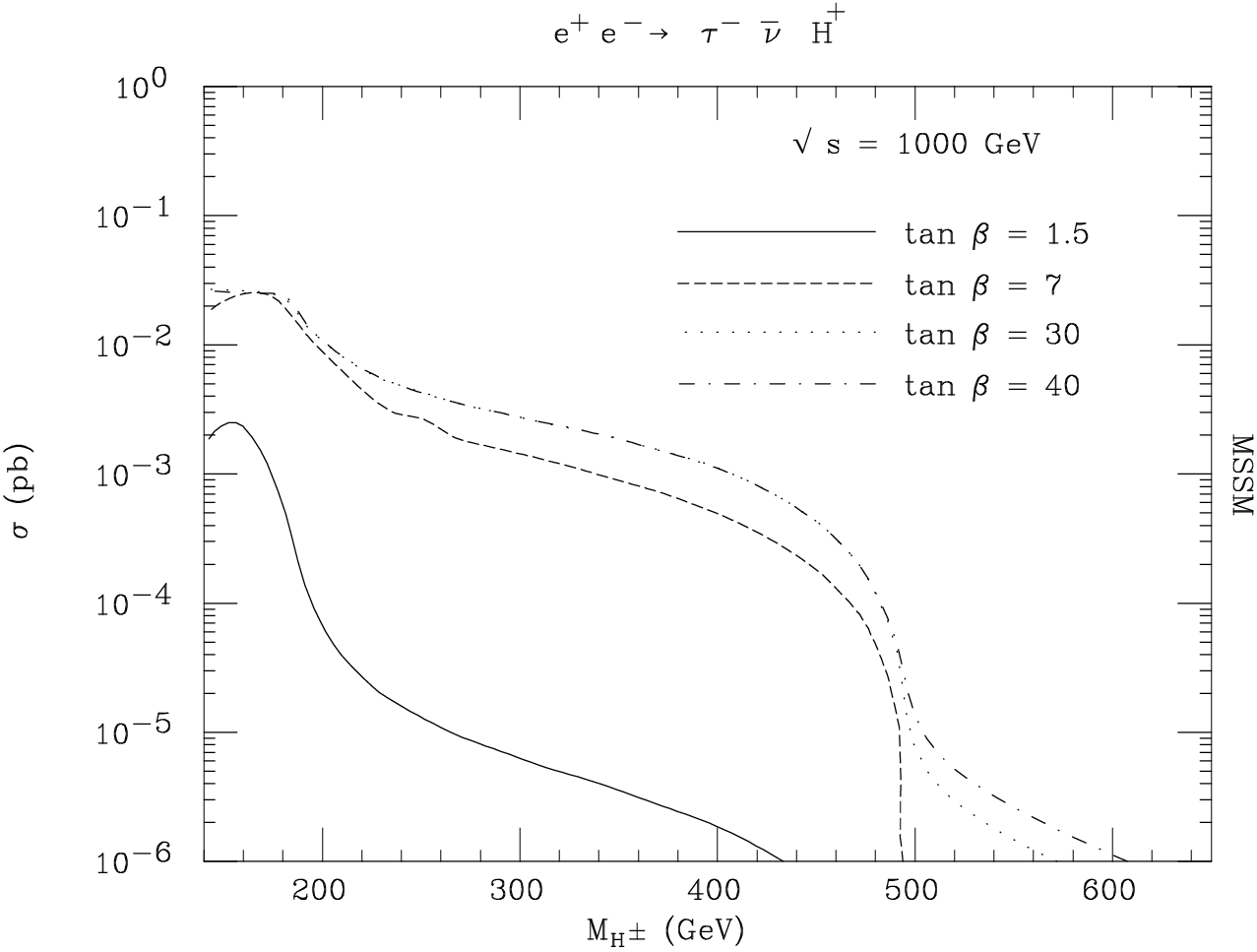,width=5cm,angle=90}\\
  \caption{Total cross sections for the tau 
           associated production channels (\ref{proc_tau}).}
%\end{figure}
%\begin{figure}[t]
  \epsfig{file=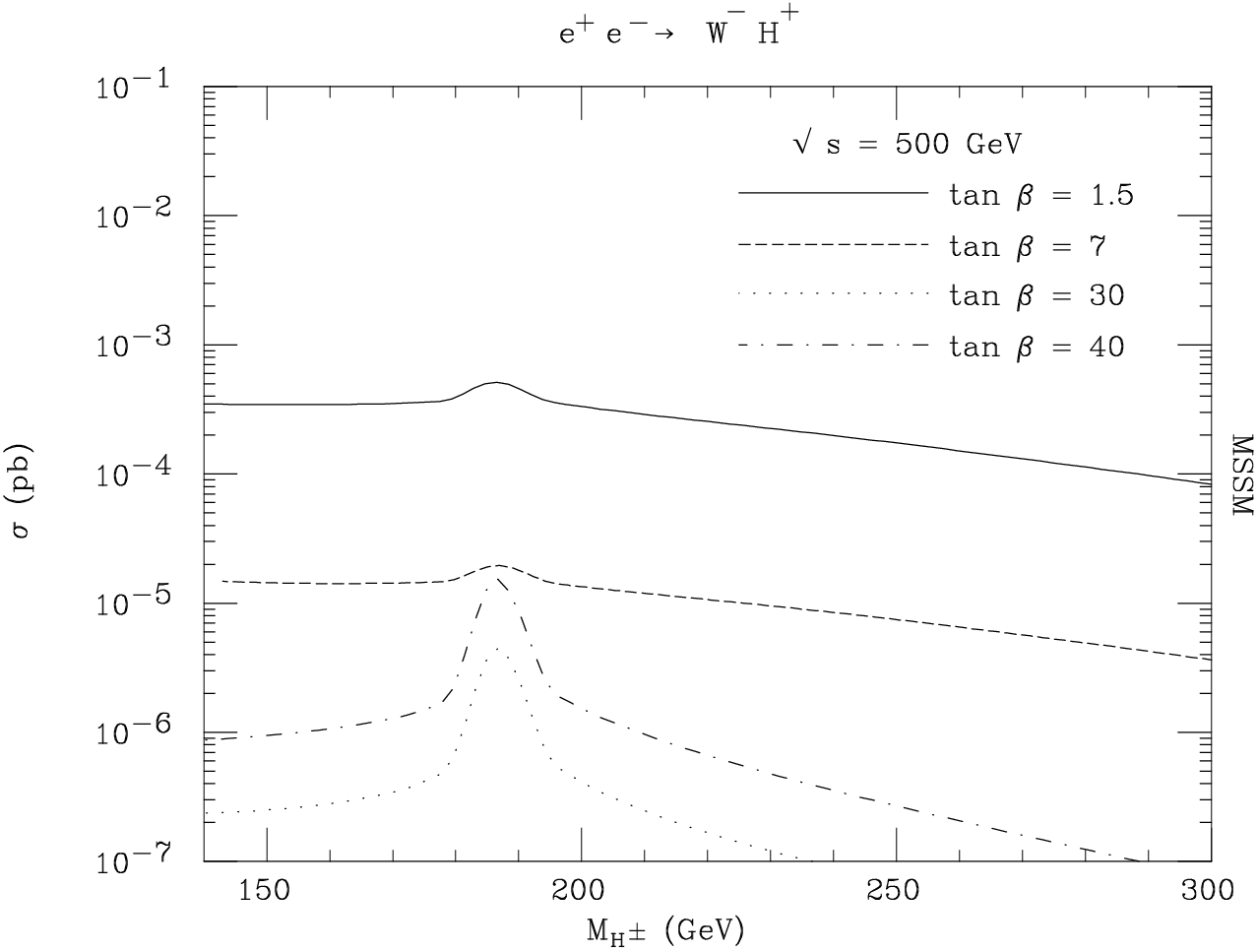,width=5cm,angle=90}
  \epsfig{file=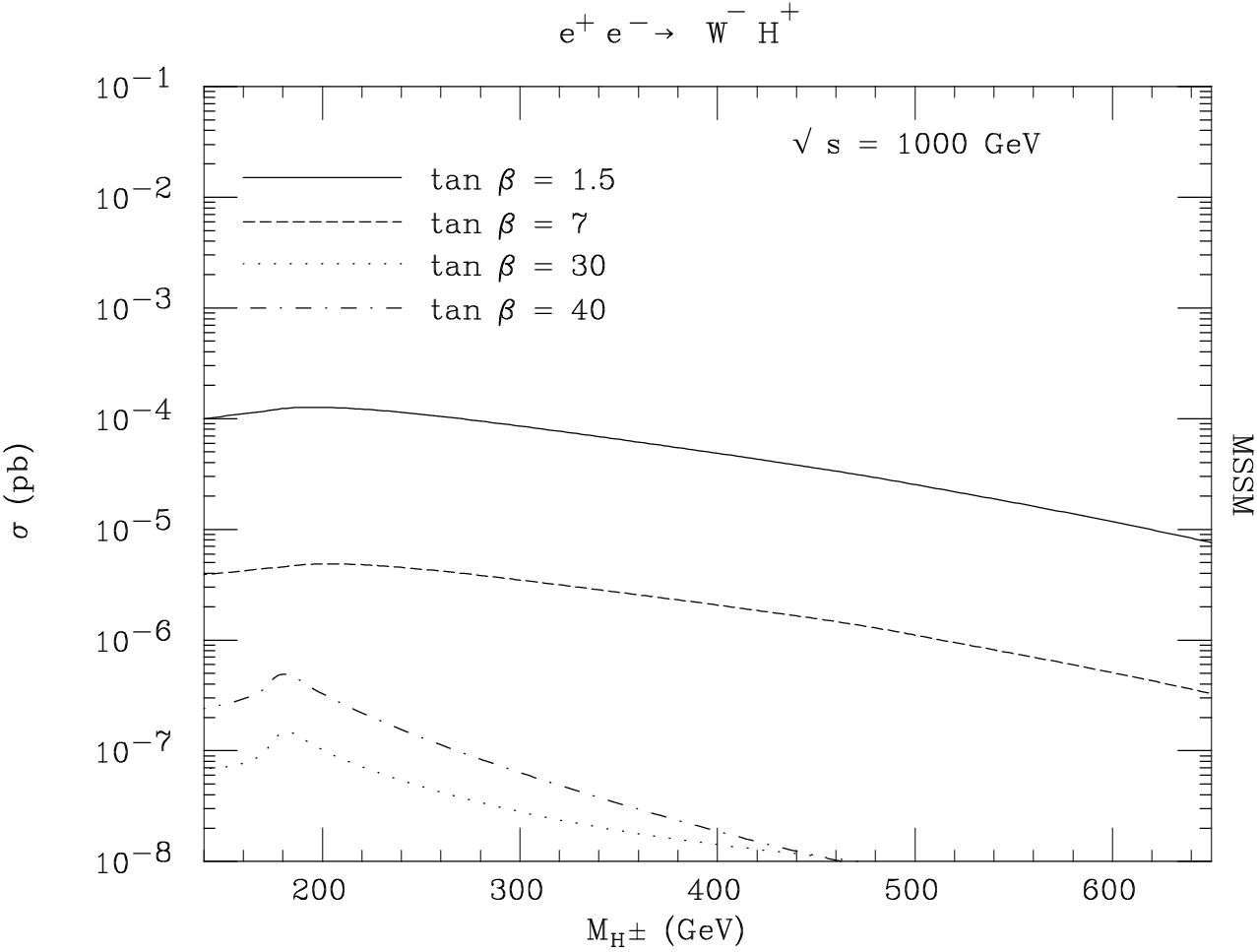,width=5cm,angle=90}\\
  \caption{Total cross sections for the $W^- H^+$ associated
  production process (\ref{proc_wh}).}
\end{figure}  

The $t \overline{b}$ (or $\overline{t} b$) associated production 
process (\ref{proc_tb}) has a similar structure to 
$\tau\nu_\tau H^\pm$ modes, but the $\tan\beta$ dependence is 
opposite: the rate is larger for smaller $\tan\beta$ values. 
The cross section is suppressed above the pair production 
threshold in comparison with that of $\tau\nu_\tau H^\pm$ modes 
because of the difference in the phase space.  

Figure~2 shows the cross section of the one-loop induced processes 
$e^-e^+ \to W^\pm H^\mp$ (\ref{proc_wh}). For small $\tan\beta$ values, 
the $H^\pm t b$ coupling included in each quark-loop diagram is large 
so that the cross section becomes substantial: the $\tan\beta$ dependence 
of the cross section is $\sim m_t^4\cot^2\beta$ at small $\tan\beta$ 
and $\sim m_b^4\tan^2\beta$ at very large $\tan\beta$. 
The cross section remains large beyond the pair production kinematic 
limit. When the dominant $H^\pm$ decay mode is $t\bar b$, we can deal 
with the top pair background by reconstructing the final state 
and eliminating events with two top quarks. This is expected to reduce the 
top pair background by about $\alpha_{\rm EW}$, which should enable the 
observation of the peak at the charged Higgs mass in some regions of the 
parameter space. 

For the rest of the processes, (\ref{proc_enu})-(\ref{proc_enah}), 
there is a variety of structures observed in the cross sections. 
Their rates, however, turned out to be very small for larger 
$M^{}_{H^\pm}$ 
than the pair production threshold. This is owing to the reasons 
such as coupling suppression, cancellation between Feynman diagrams, 
and/or smaller phase space. 
Details of the numerical results are shown in Ref.\cite{kmo}. 
As mentioned, the processes (\ref{proc_tb}), (\ref{proc_zwh}), 
(\ref{proc_lwh}), (\ref{proc_bwh}) and (\ref{proc_awh})
lead to the final state $b\bar bH^\pm W^\mp$. 
Some channels have large cross sections for 
$M^{}_{H^\pm} < \sqrt{s}/2$\cite{kmo}. 
The variety of resonance structures imply little interference 
among these processes. 
The $e^-e^+\to b\bar bH^\pm W^\mp$ modes, therefore, complement 
the pair production in the charged Higgs study at LC especially 
below the threshold of the pair production\cite{ee_bbWH}. 
The procedure for the tagging of the final state 
$b\bar bH^\pm W^\mp$ is outlined in Ref.\cite{kmo}.

In either process, information of the tau\cite{tau_polarisation} 
and the top\cite{top_polarisation} polarization 
can be used to help identify the charged Higgs boson.

\section*{Conclusion}

We found that the $\tau\bar\nu_\tau H^+$ channel and the loop induced 
$H^\pm W^\mp$ channel are the most promising channels for studying 
charged Higgs phenomenology beyond the kinematic limit for pair production. 
The $H^\pm W^\mp$ channel is enhanced at low $\tan\beta$,
whereas the $\tau\bar\nu_\tau H^+$ channel is enhanced at 
large $\tan\beta$.
For the charged Higgs bosons whose mass is below the pair production
threshold, some of the single production channels which lead to the 
$b\bar bH^\pm W^\mp$ final state complement the pair production. 
Some of their cross sections become large, but they are rapidly 
suppressed above the kinematic limit of the pair production. 
Our results motivate further study of the decay modes, the hadronic 
final states and the backgrounds relevant to these charged Higgs boson 
single production processes.

 \end{document}